\documentclass[journal,twoside,web]{ieeecolor}
\usepackage{tmi}
\usepackage{cite}
\usepackage{amsmath,amssymb,amsfonts}
\usepackage{algorithmic}
\usepackage{graphicx}
\usepackage{textcomp}
\usepackage{booktabs}
\usepackage{bm}
\usepackage{xcolor}
\usepackage{multirow}
\usepackage{url}
\usepackage{hyperref}
\usepackage{dblfloatfix}
\hypersetup{hidelinks=true}
\DeclareRobustCommand{\capred}[1]{\texorpdfstring{\textcolor{red}{#1}}{#1}}
\DeclareRobustCommand{\capblue}[1]{\texorpdfstring{\textcolor{blue}{#1}}{#1}}
\def\BibTeX{{\rm B\kern-.05em{\sc i\kern-.025em b}\kern-.08em
    T\kern-.1667em\lower.7ex\hbox{E}\kern-.125emX}}
\makeatletter
\def\@journal{}
\def\@pubid{}
\def\@doi{}
\makeatother

\begin{document}

\title{A deep dictionary network-based foundation model for ultra-low-dose CT denoising}

\author{
	Baoshun Shi, Shuangyi Yang, Ke Jiang, Bin Zhu, Zhanli Hu, and Huazhu Fu
	\thanks{This work was supported by the National Natural Science Foundation of China under Grant No. 62371414, by the Hebei Natural Science Foundation under Grant No. F2025203070, by the Beijing Natural Science Foundation--Haidian Original Innovation Joint Fund, Key Research Program under Grant L242067, by the General Open Fund Project of State Key Laboratory of Medical Imaging Science and Technology Systems, and by the Scientific Research Cultivation Project (Science and Engineering)-Basic Innovation Research Cultivation Project (Science and Engineering, Post-2021) under Grant No. 2025LGZD002. (Corresponding author: Baoshun Shi, e-mail: shibaoshun@ysu.edu.cn.)}
	\thanks{Baoshun Shi, Shuangyi Yang, and Ke Jiang are with the School of Information Science and Engineering, Yanshan University, Qinhuangdao 066004, Hebei, China, and also with the Hebei Key Laboratory of Information Transmission and Signal Processing, Yanshan University, Qinhuangdao 066004, Hebei, China.}
	\thanks{Bin Zhu is with the Department of Orthopedics, Beijing Friendship Hospital, Capital Medical University, Beijing, China.}
	\thanks{Zhanli Hu is with the Lauterbur Research Center for Biomedical Imaging, Shenzhen Institute of Advanced Technology, Chinese Academy of Sciences, Shenzhen 518055, China.}
	\thanks{Huazhu Fu is with the Institute of Advanced Intelligence and Computing (IAIC), Agency for Science, Technology and Research (A*STAR), Singapore 138632.}
}

\maketitle
\begin{abstract}
	Ultra-low-dose computed tomography (ULDCT) reduces radiation exposure but suffers from severe noise that degrades diagnostic image quality. Existing deep learning-based denoising methods are typically trained in an organ-specific fashion, resulting in limited generalization across heterogeneous multi-organ imaging scenarios. Foundation models present a promising all-in-one paradigm for unified multi-organ denoising. However, their architectures suffer from poor interpretability and rely on heuristic training strategies. To address these limitations, we propose an architecture-interpretable foundation model based on the deep dictionary network (DDN) for unified multi-organ ULDCT denoising. Inspired by multilayer sparse representation theory, DDN cascades convolutional sparse coding layers with iterative soft-thresholding, providing inherent architectural interpretability. Furthermore, a dynamic dictionary module and a threshold generation module are embedded within each layer to enhance representation ability. We conduct DDN pre-training on more than one million multi-organ normal-dose CT images by recovering clean images from Gaussian-noised inputs. Sparse regularization is additionally imposed on latent feature representations, guiding the network to learn compact and noise-robust priors. The complete architecture is jointly fine-tuned on multi-organ ULDCT datasets, enabling a single unified model to perform denoising across diverse anatomical regions. Extensive experiments validate that our proposed method achieves state-of-the-art performance and consistently surpasses competing ULDCT methods across all multi-organ benchmarks under the few-shot learning setting.
\end{abstract}


\begin{figure*}[!t]
	\centering
	\includegraphics[
	width=\textwidth,
	trim=8mm 8mm 14mm 8mm,
	clip
	]{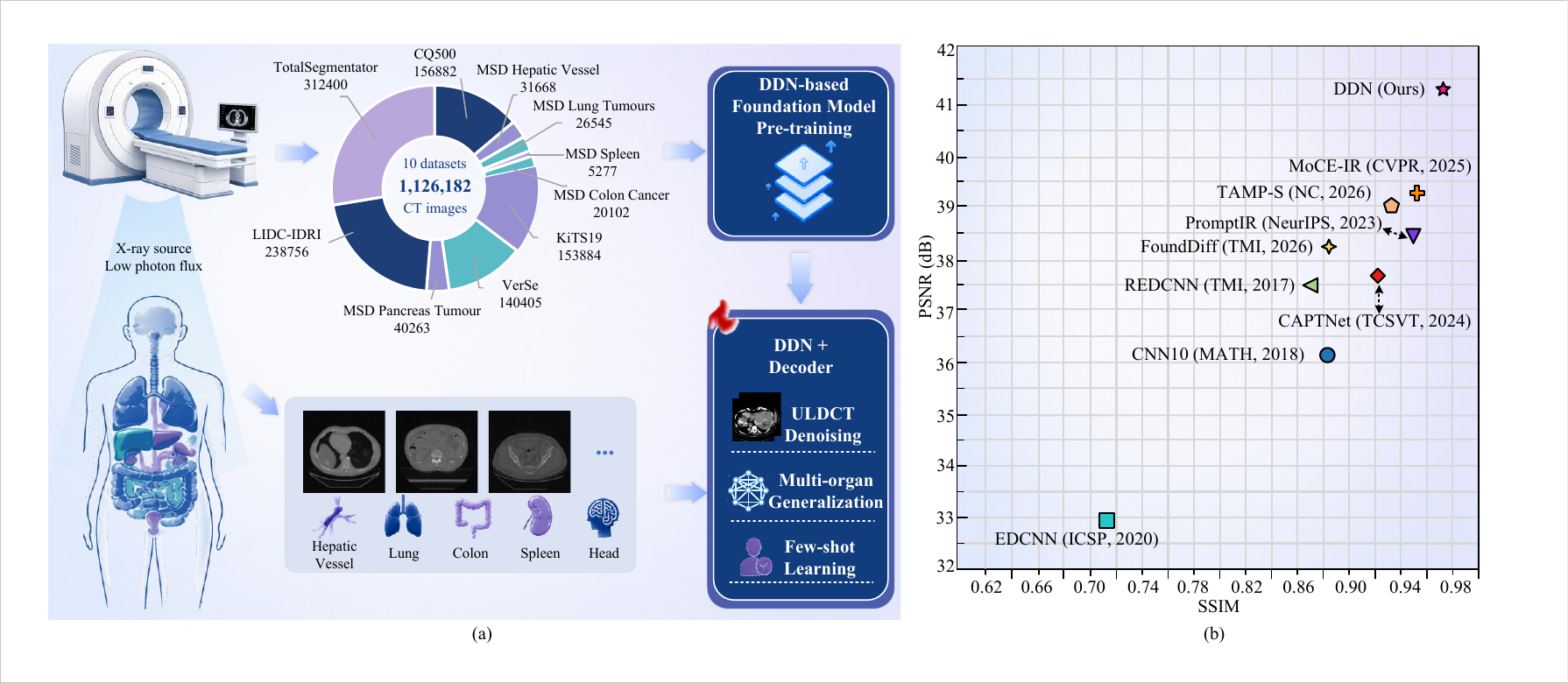}
	\caption{Overview of the proposed DDN-based pre-training framework and few-shot denoising performance. (a) DDN is pretrained on more than one million multi-organ CT images from ten datasets and then combined with a decoder for multi-organ ULDCT denoising, cross-organ generalization, and few-shot learning. (b) Average PSNR and SSIM comparison of different methods on test datasets under the few-shot setting, where only 20 training images are used for each organ. DDN achieves the highest PSNR and SSIM among the compared methods, indicating better denoising performance under limited training data.}
	\label{fig:fewshot_overview}
\end{figure*}

\section{Introduction}
\label{sec:introduction}
\IEEEPARstart{C}{omputed} tomography (CT) is an indispensable imaging modality for clinical diagnosis and disease screening. However, the associated X-ray exposure may increase the risk of cancer and other adverse health effects \cite{1}. Ultra-low-dose CT (ULDCT) can reduce radiation exposure by lowering the tube current or incident photon flux \cite{2}, but the reduced photon count introduces severe quantum noise into the projection data. After the filtered back-projection (FBP) operator, the reconstructed images often contain strong noise and streak-like artifacts that obscure fine anatomical structures and reduce diagnostic reliability. Effective ULDCT denoising methods should be developed to suppress severe noise while retaining clinically relevant details.

In recent years, deep learning-based methods have achieved promising performance in ULDCT image denoising by learning the mapping from noisy images to normal-dose CT images \cite{3,4,5}. Despite their effectiveness, most existing methods are still trained in an organ-specific manner, where a separate model is built for each anatomical region. Such a paradigm has two main limitations: storing multiple organ-specific models increases storage and deployment costs of deep neural networks (DNNs), while each model learns only from its own anatomical data, making it difficult to fully exploit general CT image priors shared across organs. As a result, the generalization capability of existing denoising methods remains limited in heterogeneous multi-organ ULDCT scenarios.

Foundation models improve the generalization of DNNs through large-scale pre-training and downstream adaptation \cite{6}. Representative visual foundation models include self-distillation approaches \cite{7} and masked image modeling methods \cite{8}. From the perspective of network architecture, most existing methods adopt ViT- or Transformer-based backbones \cite{7,8} and recent methods focus on state space model (SSM), which can effectively model long-range dependencies with linear computational complexity \cite{9}. From the perspective of application, these advances have also extended foundation models to medical applications, including computational pathology and scalable medical image encoding. However, most network architectures of existing foundation models remain difficult to interpret and primarily target generic visual representation learning or medical image understanding, rather than unified multi-organ ULDCT denoising, which requires effective noise suppression while preserving fine anatomical details.

Deep unfolding networks improve model interpretability by unrolling iterative optimization algorithms into trainable layers \cite{10,11}. However, existing unfolding networks are mostly designed for specific inverse problems, such as CT and magnetic resonance imaging reconstructions \cite{12,13}. Their task-dependent formulations limit their suitability for building generalizable foundation models. In contrast, sparse representation-based unfolding offers a more flexible image-domain prior that can describe common structural patterns shared across CT images. However, convolutional sparse coding (CSC) unfolding networks are usually derived from single-layer sparse representation, limiting their capacity to characterize the complex multi-organ anatomical structures required for unified ULDCT denoising \cite{10,14}.

To address these limitations, we propose an architecture-interpretable foundation model based on what we call the deep dictionary network (DDN). By extending single-layer sparse representation unfolding into a multilayer deep dictionary architecture, DDN combines the generalization ability of foundation models with the structural interpretability of sparse representation unfolding. A denoising objective and sparse regularization are further used during pre-training to learn compact and noise-robust CT priors from large-scale multi-organ data. The contributions are summarized as follows:

$\bullet$ We propose an architecture-interpretable foundation model based on DDN for unified multi-organ ULDCT denoising. In contrast to existing foundation models with limited architectural interpretability, DDN is derived from multilayer sparse representation theory and unfolds convolutional sparse coding into cascaded dictionary layers with iterative soft-thresholding, providing a transparent and principled backbone for CT denoising.

$\bullet$ We design a dynamic dictionary module (DDM) and a threshold generation module (TGM) to enhance DDN's representation while preserving sparse-coding interpretability. The DDM generates input-dependent dictionary atoms from image features to replace fixed kernels, while the TGM fuses features from local, large, non-local, and global sub-modules to predict spatially adaptive thresholds. This allows soft-thresholding to better handle heterogeneous anatomical structures and spatially varying noise. We also verify the effectiveness of these sub-modules within TGM through ablation studies.

$\bullet$ We develop a denoising-oriented pre-training strategy with sparse representation regularization for the DDN-based foundation model. DDN is pretrained on more than one million multi-organ normal-dose CT images by recovering clean images from Gaussian-noised inputs, while an $\ell_1$ constraint is imposed on latent sparse representation. This strategy promotes compact, noise-robust, and transferable CT denoising priors, providing effective initialization for downstream multi-organ ULDCT denoising.

$\bullet$ As depicted in Fig.~\ref{fig:fewshot_overview}, the proposed framework integrates large-scale multi-organ pre-training with downstream ULDCT denoising. Extensive experiments on synthetic multi-organ ULDCT datasets and real experimental CT data show that DDN consistently outperforms existing methods across anatomical regions, demonstrating strong cross-organ generalizability. Furthermore, under the few-shot setting with only 20 training images per organ, DDN achieves the best average PSNR and SSIM values among the compared methods, confirming the effectiveness of the pretrained DDN priors under limited training data.

The remainder of this paper is organized as follows. Section II reviews the existing ULDCT denoising methods. Section III presents the architecture of the proposed deep dictionary network. Section IV describes the proposed foundation model training strategy. Section V reports experimental results and performance analysis, and Section VI concludes the paper.

\section{Existing Ultra-Low-Dose CT Denoising Methods}
Ultra-low-dose CT (ULDCT) denoising aims to reduce severe noise and artifacts while preserving fine anatomical structures. Traditional methods commonly formulate CT denoising as a regularization problem with handcrafted regularizers. Although these methods can improve image quality, they often require careful parameter tuning and suffer from high computational cost, which limits their practical efficiency.

With the development of deep learning, learning-based methods have become dominant solutions for ULDCT denoising. According to the domain in which the network is applied, existing methods can be roughly divided into sinogram-domain, image-domain, and dual-domain approaches. Sinogram-domain methods directly process noisy projection data before image reconstruction \cite{4}. By restoring the corrupted measurements in the projection domain, these methods can reduce noise at the data level, but secondary artifacts may still be introduced during the subsequent reconstruction process. Image-domain methods directly operate on reconstructed CT images and are therefore more flexible, since they do not require access to raw projection data or scanner-specific parameters. Representative methods include REDCNN \cite{3}, which improves denoising stability through residual encoder-decoder learning, and EDCNN \cite{5}, which enhances edge preservation with edge-aware compound losses. Dual-domain methods jointly exploit projection-domain and image-domain information to improve restoration performance. By incorporating data consistency, domain interaction, or dual-domain learning strategies, these methods can achieve stronger denoising performance \cite{15}. However, dual-domain methods usually require projection data and accurate imaging geometry, which may not always be available in practical clinical scenarios.

Most existing CT denoising methods are developed for specific anatomical regions or data distributions, resulting in limited generalization across heterogeneous multi-organ ULDCT scenarios. Unified image restoration provides a potential solution by handling multiple degradations within a single model. Early all-in-one methods distinguish degradation-specific features through contrastive learning \cite{16}, while subsequent approaches employ prompt mechanisms \cite{17,18,19} for adaptive restoration. PromptCT \cite{20} addresses multiple sparse-view settings, while FoundDiff \cite{21} explores unified LDCT denoising across doses and anatomical regions. However, existing methods remain limited in degradation types, sampling settings, or anatomical coverage, whereas cross-organ generalization in broader multi-organ ULDCT scenarios remains underexplored. Developing a unified and generalizable model for multi-organ ULDCT denoising is still an important and challenging problem.

\begin{figure*}[!t]
	\centering
	\includegraphics[
	width=\textwidth,
	trim=8mm 7mm 5mm 8mm,
	clip
	]{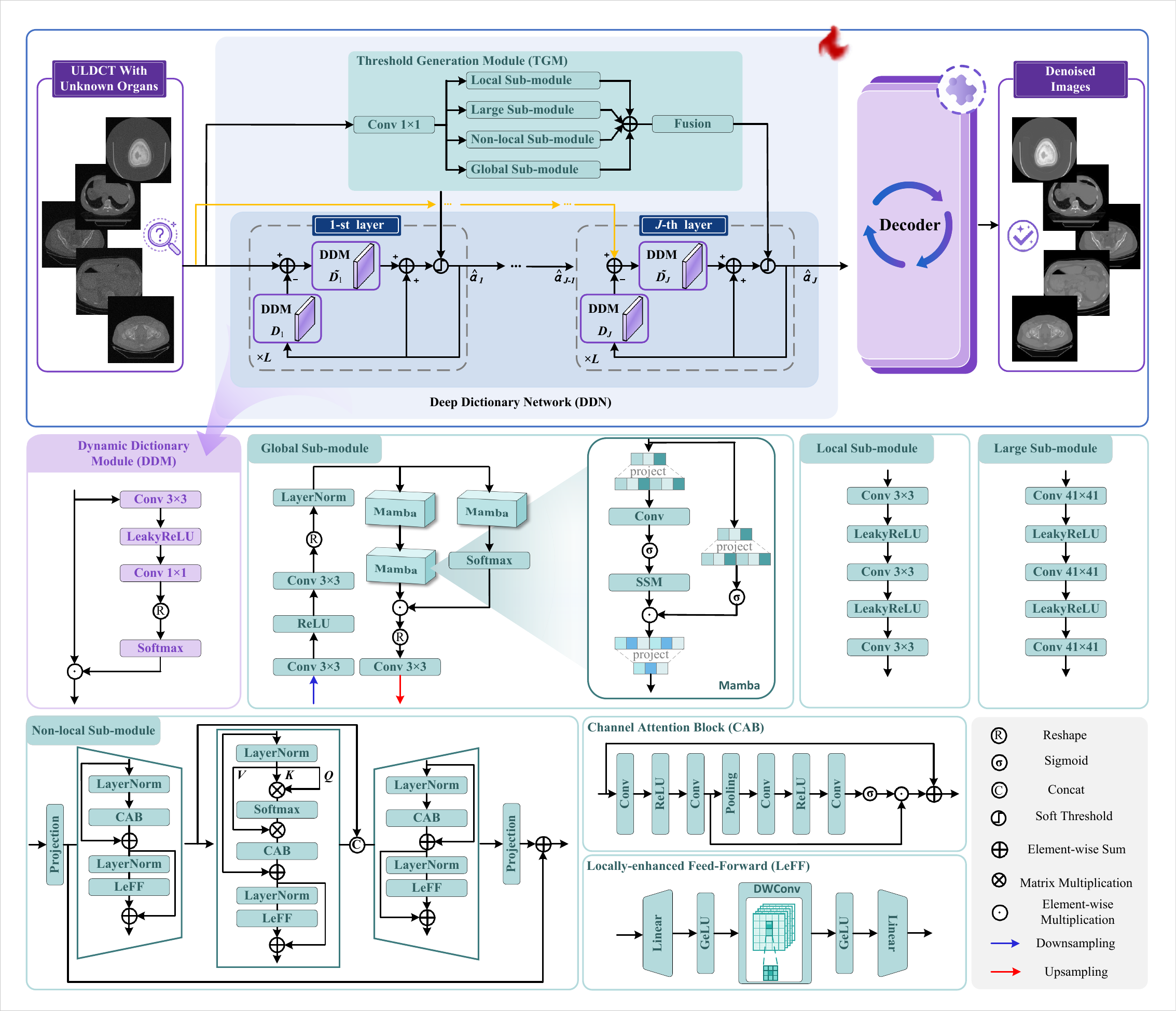}
	\caption{Architecture of the proposed DDN. DDN cascades convolutional sparse coding layers with iterative updates, where each layer incorporates input-dependent DDM ($\bm{D}_j$ and $\tilde{\bm{D}}_j$) and adaptive soft-thresholding. The TGM integrates local, large, non-local, and global features to produce a spatially adaptive threshold map shared across the iterative steps within each layer.}
	\label{fig:DDN}
\end{figure*}

\section{The Proposed Deep Dictionary Network}
\normalcolor
\subsection{Single-layer sparse coding}
Given an input $\bm{x}\in\mathbb{R}^{N}$ and an overcomplete dictionary $\bm{D}\in\mathbb{R}^{N\times M}$ ($M>N$), sparse coding aims to seek a sparse representation $\bm{\alpha}\in\mathbb{R}^{M}$ such that $\bm{D}\bm{\alpha}\approx\bm{x}$. This problem is formulated as the following $\ell_1$-regularized optimization problem \cite{14}:
\begin{equation}
	\min_{\bm{\alpha}}
	\frac{1}{2}\|\bm{x}-\bm{D}\bm{\alpha}\|_2^2
	+\lambda\|\bm{\alpha}\|_1
	\label{eq:sparse_coding}
\end{equation}
where the two terms measure the approximation error and promote sparsity, and $\lambda$ balances data fidelity and sparsity.

The optimization problem in Eq.~\eqref{eq:sparse_coding} is solved using the iterative shrinkage-thresholding algorithm (ISTA) \cite{10} through alternating gradient descent and proximal mapping steps.

\noindent\textbf{First Step: Gradient Descent.}
For the smooth data fidelity term
$f(\bm{\alpha})=\frac{1}{2}\|\bm{x}-\bm{D}\bm{\alpha}\|_2^2$,
the gradient and step-size condition are
\begin{equation}
	\nabla_{\bm{\alpha}}f(\bm{\alpha}^{(l)})
	=
	-\bm{D}^{\mathrm{T}}
	(\bm{x}-\bm{D}\bm{\alpha}^{(l)}),
	\
	c\geq\sigma_{\max}(\bm{D}^{\mathrm{T}}\bm{D})
	\label{eq:gradient_step}
\end{equation}
where $c$ is an upper bound on the maximum eigenvalue of
$\bm{D}^{\mathrm{T}}\bm{D}$ and $1/c$ is the step size. The intermediate estimate is
\begin{equation}
	\bm{\alpha}^{\left(l+\tfrac{1}{2}\right)}
	=
	\bm{\alpha}^{(l)}
	+
	\frac{1}{c}\bm{D}^{\mathrm{T}}
	(\bm{x}-\bm{D}\bm{\alpha}^{(l)})
	\label{eq:grad_descent}
\end{equation}

\noindent\textbf{Second Step: Proximal Mapping.}
The proximal operator of the $\ell_1$ norm finds a sparse solution close to $\bm{\alpha}^{\left(l+\tfrac{1}{2}\right)}$:
\begin{equation}
	\bm{\alpha}^{(l+1)}
	=
	\arg\min_{\bm{\alpha}}
	\left\{
	\tau\|\bm{\alpha}\|_1
	+
	\frac{1}{2}
	\|\bm{\alpha}-\bm{\alpha}^{\left(l+\tfrac{1}{2}\right)}\|_2^2
	\right\}
	\label{eq:prox_def}
\end{equation}
where $\tau=\lambda/c$. Its element-wise solution is
$\bm{\alpha}^{(l+1)}
=\mathcal{S}_{\tau}(\bm{\alpha}^{\left(l+\tfrac{1}{2}\right)})$,
where
$\mathcal{S}_{\tau}(x)
=\operatorname{sign}(x)\cdot\max(|x|-\tau,0)$
is the soft-thresholding operator. Coefficients below $\tau$ are set to zero, while larger coefficients are shrunk by $\tau$.

\subsection{From single-layer sparse representation to deep sparse representation with dynamic dictionary}
To improve the representation capacity of conventional single-layer sparse coding, we cascade multiple dictionary layers to construct hierarchical sparse representations and replace static kernels with input-dependent dynamic dictionaries.

Motivated by multilayer sparse representation theory, we introduce $J$ cascaded convolutional sparse coding layers. The first-layer representation satisfies
$\bm{x}\approx\bm{D}_1\bm{\alpha}_1$ and $\bm{\alpha}_1$ is further represented as $\bm{\alpha}_1\approx\bm{D}_2\bm{\alpha}_2$. Extending this process gives
$\bm{x}\approx\bm{D}_1\bm{D}_2\cdots\bm{D}_J\bm{\alpha}_J$,
where $\bm{\alpha}_J$ denotes the final-layer sparse representation. In the deep unfolding framework, the dictionaries and related parameters are learned through backpropagation. Given these dictionaries, the deep sparse coding problem can be formulated as
\begin{equation}
	\begin{aligned}
		\min_{\{\bm{\alpha}_j\}_{j=1}^{J}}
		\quad&
		\frac{1}{2}
		\left\|
		\bm{x}
		-
		\bm{D}_1\bm{D}_2\cdots
		\bm{D}_J\bm{\alpha}_J
		\right\|_2^2
		+
		\lambda_j\sum_{j=1}^{J}\|\bm{\alpha}_j\|_1,
		\\
		\mathrm{s.t.}\quad&
		\forall j,\ 
		\bm{\alpha}_j
		=
		\bm{D}_{j+1}\bm{\alpha}_{j+1}.
	\end{aligned}
\end{equation}
The above formulation is a complex constrained problem. To approximately solve the problem, we proceed iteratively as follows:
\begin{equation}
	\left\{
	\begin{aligned}
		\hat{\bm{\alpha}}_1
		&=
		\underset{\bm{\alpha}_1}{\operatorname*{arg\,min}}
		\frac{1}{2}
		\left\|
		\bm{x}-\bm{D}_1\bm{\alpha}_1
		\right\|_2^2
		+
		\lambda_1\|\bm{\alpha}_1\|_1,
		\\
		\hat{\bm{\alpha}}_2
		&=
		\underset{\bm{\alpha}_2}{\operatorname*{arg\,min}}
		\frac{1}{2}
		\left\|
		\hat{\bm{\alpha}}_1-\bm{D}_2\bm{\alpha}_2
		\right\|_2^2
		+
		\lambda_2\|\bm{\alpha}_2\|_1,
		\\
		&\ \vdots
		\\
		\hat{\bm{\alpha}}_J
		&=
		\underset{\bm{\alpha}_J}{\operatorname*{arg\,min}}
		\frac{1}{2}
		\left\|
		\hat{\bm{\alpha}}_{J-1}
		-
		\bm{D}_J\bm{\alpha}_J
		\right\|_2^2
		+
		\lambda_J\|\bm{\alpha}_J\|_1.
	\end{aligned}
	\right.
	\label{eq:layerwise_sparse}
\end{equation}

Each subproblem defined above is solved using ISTA. For the $j$-th subproblem, the $(l+1)$-th iteration is given by
\begin{equation}
	\bm{\alpha}_j^{l+1}
	=
	\mathcal{S}_{\tau}
	\left(
	\bm{\alpha}_j^{l}
	+
	\frac{1}{c}\bm{D}_j^{\mathrm{T}}
	\left(
	\bm{x}
	-
	\bm{D}_j\bm{\alpha}_j^{l}
	\right)
	\right)
	\label{eq:deep_ista}
\end{equation}
where $\bm{\alpha}_j^{l}$ is the representation at iteration $l$ of layer $j$, and each layer performs $L$ iterations.

Inspired by \cite{11}, we replace the static dictionary with an input-dependent dynamic dictionary module (DDM) to improve its adaptability to heterogeneous anatomical structures. At the $j$-th layer, the dynamic dictionary is generated from the input feature as
$\bm{D}_j(\bm{x})=\mathcal{F}_j(\bm{x})$,
where $\mathcal{F}_j(\cdot)$ denotes the procedure of the DDM. This allows the dictionary atoms to adapt to the image content and enhances the representation capacity of each dictionary layer.

Accordingly, the layer-wise sparse coding is
\begin{equation}
	\bm{\alpha}_j^{l+1}
	=
	\mathcal{S}_{\tau}
	\left(
	\bm{\alpha}_j^{l}
	+
	\frac{1}{c}
	\tilde{\bm{D}}_j
	\left(
	\bm{x}
	-
	\bm{D}_j(\bm{x})\bm{\alpha}_j^{l}
	\right)
	\right)
	\label{eq:deep_ista_dynamic}
\end{equation}
To further improve representation capacity, $(\bm{D}_j(\bm{x}))^{\mathrm{T}}$ is replaced by
$\tilde{\bm{D}}_j$, which is generated by the dynamic dictionary generation module $\mathcal{F}_j(\cdot)$. The overall architecture is shown in Fig.~\ref{fig:DDN}.

\subsection{Threshold generation module}

The threshold $\tau$ in the soft-thresholding operator controls sparsity. Conventional ISTA adopts a globally fixed scalar threshold \cite{10}, which cannot adapt to heterogeneous anatomical structures and spatially varying ULDCT noise. We therefore design a module with multiple sub-modules to predict a spatially adaptive threshold map $\bm{\tau}$ from the initial coefficient estimate $\tilde{\bm{\alpha}}$, obtained by projecting the input image followed by an initial DDM. After a $1\times1$ convolutional projection, the feature map $\bm{z}$ is fed into four parallel sub-modules to extract features.

\subsubsection{Local and large sub-modules}
Since ULDCT images contain both fine details and large-scale structures, a single receptive field is insufficient for threshold generation. The local sub-module uses three successive $3\times3$ convolutions with LeakyReLU activations to capture edges and fine details, producing the local feature output $f_{\mathrm{local}}(\bm{z})$. In contrast, the large sub-module employs $41\times41$ kernels to capture broader context and large-scale anatomical structures, yielding the large-scale feature output $f_{\mathrm{large}}(\bm{z})$. These two sub-modules provide complementary information for threshold generation.

\subsubsection{Non-local sub-module}
Convolutional operations have limited ability to capture correlations between distant anatomical regions. To address this limitation, the non-local sub-module adopts Transformer to model long-range dependencies through window-based self-attention and shifted-window interaction. The projected sparse feature $\bm{z}$ is embedded as $\bm{z}_0$ and processed by $N$ channel attention (CA) transformer blocks, each consisting of a channel attention block (CAB) and a locally-enhanced feed-forward network (LeFF):
\begin{align}
	\tilde{\bm{z}}_1
	&=
	\mathrm{CAB}(\mathrm{LN}(\bm{z}_0))
	+
	\bm{z}_0,
	\\
	\bm{z}_1
	&=
	\mathrm{LeFF}(\mathrm{LN}(\tilde{\bm{z}}_1))
	+
	\tilde{\bm{z}}_1.
\end{align}
where $\mathrm{LN}(\cdot)$ denotes layer normalization.

The encoder feature is then fed into the shadow-interaction module (SIM), where shifted $w\times w$ windows enable cross-window information exchange. The window-based self-attention is formulated as
\begin{equation}
	\mathrm{Attn}(\bm{Q},\bm{K},\bm{V})
	=
	\mathrm{Softmax}
	\left(
	\frac{\bm{Q}\bm{K}^{\top}}{\sqrt{d}}+\bm{B}
	\right)\bm{V}
\end{equation}
where $\bm{Q}$, $\bm{K}$, and $\bm{V}$ are the query, key, and value features, $d$ is the head dimension, $\bm{B}$ is the relative position bias, and $\mathrm{Softmax}(\cdot)$ normalizes the attention scores into probability distributions.

Finally, the bottleneck feature, denoted as $\bm{z}_2$, is concatenated with the encoder feature $\bm{z}_1$ and fused by the decoder CA transformer blocks, producing the non-local feature output $f_{\mathrm{non-local}}(\bm{z})$:
\begin{equation}
	f_{\mathrm{non-local}}(\bm{z})
	=
	\mathrm{OutputProj}
	\bigl(
	\mathrm{Decoder}([\bm{z}_2,\bm{z}_1])
	\bigr)
	+
	\bm{z}
\end{equation}
where $\mathrm{Decoder}(\cdot)$ performs feature fusion and $\mathrm{OutputProj}(\cdot)$ produces the output. 

\subsubsection{Global sub-module}
Threshold generation also depends on the overall anatomical layout and noise distribution, which cannot be fully captured by convolutional or window-based operations. Inspired by Mamba \cite{9}, the global sub-module employs a U-shaped vision Mamba block (UVMB) to efficiently model global dependencies. The input feature $\bm{z}$ is downsampled to $32\times32$, processed by two $3\times3$ convolutional layers, and flattened into a normalized sequence $\tilde{\bm{z}}$. UVMB uses independently parameterized Mamba blocks in a representation path and a weighting path to capture global dependencies and emphasize informative responses. The representation path refines the global representation $\bm{r}$ and the weighting path generates adaptive weights $\bm{\omega}$ to modulate feature importance. Their element-wise product is then projected and upsampled:
\begin{equation}
	f_{\mathrm{global}}(\bm{z})
	=
	\mathrm{Upsample}
	\bigl(
	\mathrm{Conv}(\bm{\omega}\odot\bm{r})
	\bigr)
\end{equation}
where $\odot$ denotes element-wise multiplication, and $f_{\mathrm{global}}(\bm{z})$ represents the global feature output.

The four sub-modules provide complementary information, and their summed outputs are fed into a fusion sub-module consisting of a $1\times1$ convolution and a sigmoid function to produce the spatially adaptive threshold map:
\begin{equation}
	\bm{\tau} =\sigma\Bigl(\mathrm{Conv}\bigl(
	f_{\mathrm{local}}(\bm{z})+
	f_{\mathrm{large}}(\bm{z})+
	f_{\mathrm{non-local}}(\bm{z})+
	f_{\mathrm{global}}(\bm{z})\bigr)\Bigr)
\end{equation}
The resulting threshold map is shared across the soft-thresholding operations within the corresponding dictionary layer.

\subsection{Overall architecture of the proposed DDN}
In summary, the above sparse coding iterative process is unfolded into the DDN by cascading $J$ dictionary layers, each performing $L$ iterative updates. As shown in Fig.~\ref{fig:DDN}, an input CT image is projected into the feature space and progressively refined into hierarchical sparse representations. Within each layer, the static dictionary and fixed threshold of conventional ISTA are replaced by the input-dependent DDM and TGM. The threshold map integrates local, large, non-local, and global feature information. Consequently, DDN combines the inherent interpretability of multilayer sparse representations with improved adaptability and representation capacity for unified multi-organ ULDCT denoising.

\section{The Proposed Foundation Model Training Strategy}
This section describes the pre-training strategy for the DDN-based foundation model. DDN is pretrained on more than one million multi-organ normal-dose CT images by recovering clean images from Gaussian-noised inputs, aligning the pre-training objective with ULDCT denoising.

Let $\bm{u}$ and $\bm{y}$ denote a clean CT image normalized to $[0,1]$ and its noisy counterpart, respectively. Through cascaded sparse coding, DDN maps $\bm{y}$ to the final latent sparse representation
$\bm{\alpha}_{J}=F_{\mathrm{DDN}}(\bm{y})$. 
Following the sparse coding formulation in Section~III, the denoising pre-training problem is formulated as:
\begin{equation}
	\min_{\bm{\alpha}_{J}}
	\frac{1}{2}
	\left\|
	\mathcal{P}(\bm{\alpha}_{J})-\bm{u}
	\right\|_2^2
	+
	\beta
	\left\|
	\bm{\alpha}_{J}
	\right\|_1
	\label{eq:pretraining_problem}
\end{equation}
where $F_{\mathrm{DDN}}(\cdot)$ denotes the denoising process of the proposed DDN, and $\mathcal{P}(\cdot)$ maps $\bm{\alpha}_{J}$ from the latent representation space to the spatial image domain.

During pre-training, Gaussian noise is added to the clean image to generate the noisy input:
\begin{equation}
	\bm{y}=\bm{u}+\bm{n}, \quad
	\bm{n}\sim
	\mathcal{N}
	\left(
	0,
	\left(\frac{\sigma}{255}\right)^2\bm{I}
	\right)
	\label{eq:Gaussian_noise}
\end{equation}
where $\bm{n}$ is Gaussian noise, $\sigma$ controls the noise level, and $\bm{I}$ is the identity matrix. During pre-training, $\sigma$ can be fixed or randomly sampled from a predefined range.

To constrain the denoised image to approach the clean target, the mean squared error is adopted as the denoising loss, i.e.,
$\mathcal{L}_{\mathrm{de}}
=
\frac{1}{|\Omega|}
\left\|
\hat{\bm{u}}-\bm{u}
\right\|_2^2$, where $\hat{\bm{u}}$ denotes the denoised image and $|\Omega|$ is the number of image pixels. This loss measures the pixel-wise discrepancy between $\hat{\bm{u}}$ and $\bm{u}$, guiding DDN to recover clean CT images.

Although the denoising loss constrains the output image, it does not explicitly regularize the sparse coefficients produced by DDN. Since $\bm{\alpha}_{J}$ carries explicit sparse coding meaning, we impose
$\mathcal{L}_{\mathrm{sparse}}=\|\bm{\alpha}_{J}\|_1$
to encourage a compact sparse representation during pre-training.

The overall pre-training objective combines the denoising loss and sparse regularization:
\begin{equation}
	\begin{aligned}
		\mathcal{L}
		&=
		\mathcal{L}_{\mathrm{de}}
		+
		\beta\mathcal{L}_{\mathrm{sparse}} \\
		&=
		\frac{1}{|\Omega|}
		\left\|
		\mathcal{P}\left(F_{\mathrm{DDN}}(\bm{y})\right)-\bm{u}
		\right\|_2^2
		+
		\beta
		\left\|
		F_{\mathrm{DDN}}(\bm{y})
		\right\|_1
	\end{aligned}
	\label{eq:pretraining_loss}
\end{equation}
where $\beta$ is the weighting coefficient.

In summary, the proposed pre-training strategy learns Gaussian-noised denoising priors from large-scale multi-organ CT images by jointly optimizing image denoising loss and sparse representation regularization. Consequently, the pretrained DDN-based foundation model learns compact and noise-robust representations and provides effective initialization for downstream multi-organ ULDCT denoising.

\begin{table*}[!t]
	\begingroup
	\centering
	\caption[Quantitative evaluations for different ULDCT denoising methods on the test datasets.]%
	{\textnormal{Quantitative evaluations for different ULDCT denoising methods on the test datasets. We report the average [PSNR (dB)$\uparrow$/SSIM$\uparrow$/RMSE$\downarrow$] values for each organ. The best results are highlighted in \capred{red} and the second-best results are \capblue{blue}.}}
	\label{tab:comparison}
	
	\footnotesize
	\setlength{\tabcolsep}{2.6pt}
	\renewcommand{\arraystretch}{1.10}
	\setlength{\heavyrulewidth}{0.8pt}
	\setlength{\lightrulewidth}{0.5pt}
	\setlength{\cmidrulewidth}{0.5pt}
	
	\begin{tabular*}{\textwidth}{@{\extracolsep{\fill}}lcccccc@{}}
		\toprule
		Method & Head & Colon & Hepatic Vessel & Lung & Spleen & Average \\
		\midrule
		
		\multicolumn{7}{c}{\textnormal{LDCT Models}} \\
		\midrule
		
		FBP
		& 28.02/0.8838/0.0398
		& 31.30/0.8836/0.0272
		& 30.98/0.8775/0.0284
		& 31.30/0.8717/0.0273
		& 30.76/0.8677/0.0290
		& 30.47/0.8769/0.0303 \\
		
		REDCNN \cite{3}
		& 41.31/0.9722/0.0086
		& 43.43/0.9705/0.0068
		& 40.24/0.9521/0.0098
		& 40.78/0.9476/0.0091
		& 40.86/0.9555/0.0091
		& 41.32/0.9596/0.0087 \\
		
		CNN10 \cite{4}
		& 41.00/0.9751/0.0090
		& 42.63/0.9751/0.0074
		& 39.69/0.9519/0.0104
		& 40.16/0.9491/0.0098
		& 39.84/0.9549/0.0102
		& 40.67/0.9612/0.0094 \\
		
		EDCNN \cite{5}
		& 39.05/0.9613/0.0112
		& 41.51/0.9599/0.0084
		& 38.51/0.9310/0.0119
		& 39.23/0.9372/0.0109
		& 38.66/0.9293/0.0117
		& 39.39/0.9437/0.0108 \\
		
		FoundDiff \cite{21}
		& 42.48/0.9837/0.0076
		& 43.42/0.9811/0.0068
		& 42.04/0.9717/0.0080
		& 41.05/0.9596/0.0089
		& 41.31/0.9702/0.0086
		& 42.06/0.9733/0.0080 \\
		
		\midrule
		\multicolumn{7}{c}{\textnormal{All-in-one Models}} \\
		\midrule
		
		PromptIR \cite{17}
		& 42.94/0.9880/0.0072
		& 43.79/0.9838/0.0065
		& 42.06/0.9734/0.0079
		& 40.76/0.9591/0.0092
		& 41.43/0.9737/0.0085
		& 42.20/0.9756/0.0079 \\
		
		CAPTNet \cite{29}
		& 41.83/0.9867/0.0082
		& 43.63/0.9831/0.0066
		& 41.94/0.9732/0.0080
		& 40.73/0.9592/0.0092
		& 41.29/0.9727/0.0086
		& 41.88/0.9750/0.0081 \\
		
		MoCE-IR \cite{30}
		& \textcolor{blue}{42.97}/\textcolor{blue}{0.9881}/\textcolor{blue}{0.0072}
		& \textcolor{blue}{44.31}/\textcolor{blue}{0.9843}/\textcolor{blue}{0.0061}
		& \textcolor{red}{42.19}/\textcolor{red}{0.9748}/\textcolor{red}{0.0078}
		& 41.09/0.9612/0.0088
		& 41.46/\textcolor{blue}{0.9739}/0.0085
		& \textcolor{blue}{42.41}/\textcolor{blue}{0.9765}/\textcolor{blue}{0.0077} \\
		
		TAMP-S \cite{6}
		& 42.65/0.9873/0.0074
		& 44.27/0.9838/0.0061
		& 42.19/\textcolor{blue}{0.9739}/0.0078
		& \textcolor{blue}{41.23}/\textcolor{blue}{0.9616}/\textcolor{blue}{0.0087}
		& \textcolor{blue}{41.47}/0.9724/\textcolor{blue}{0.0085}
		& 42.36/0.9758/0.0077 \\
		
		\textbf{DDN}
		& \textcolor{red}{44.32}/\textcolor{red}{0.9894}/\textcolor{red}{0.0062}
		& \textcolor{red}{44.81}/\textcolor{red}{0.9846}/\textcolor{red}{0.0058}
		& \textcolor{blue}{42.08}/0.9736/\textcolor{blue}{0.0079}
		& \textcolor{red}{41.72}/\textcolor{red}{0.9629}/\textcolor{red}{0.0082}
		& \textcolor{red}{42.02}/\textcolor{red}{0.9746}/\textcolor{red}{0.0079}
		& \textcolor{red}{42.99}/\textcolor{red}{0.9770}/\textcolor{red}{0.0072} \\
		
		\bottomrule
	\end{tabular*}
	
	\endgroup
	\normalcolor
\end{table*}
\normalcolor

\begin{figure*}[!t]
	\centering
	\includegraphics[
	width=\textwidth,
	]{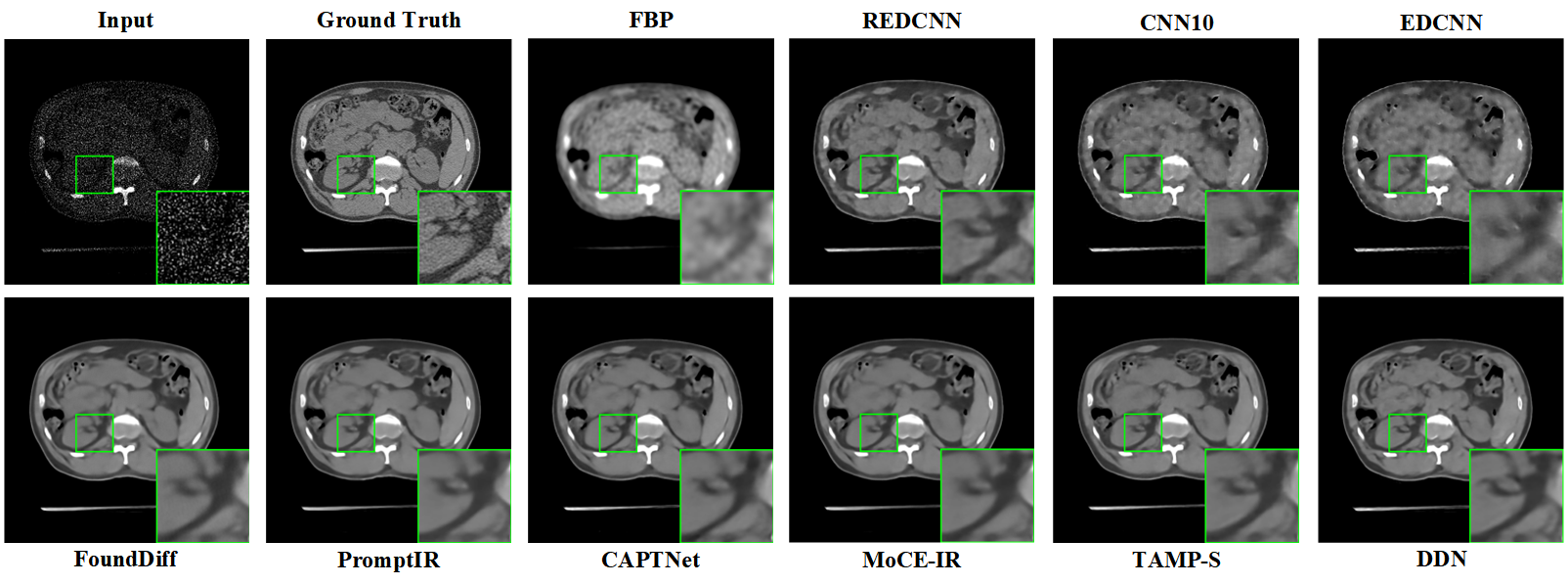}
	\caption{Visual comparison of different ULDCT denoising methods on a representative lung case from the internal datasets. Images are displayed using an HU window of [50, 500]. Green boxes indicate enlarged regions for detailed comparison of noise suppression and anatomical structure preservation.}
	\label{fig:visual_indomain}
\end{figure*}

\section{Experiments}
\label{sec:experiments}
\subsection{Datasets}
\subsubsection{Pre-training datasets}
The DDN-based foundation model is pretrained on ten publicly available multi-organ CT datasets, including CQ500 \cite{22}, five CT tasks from the medical segmentation decathlon (MSD), i.e., Hepatic Vessel, Lung Tumours, Spleen, Colon Cancer, and Pancreas Tumours \cite{23}, KiTS19 \cite{24}, VerSe \cite{25}, LIDC-IDRI \cite{26} and TotalSegmentator \cite{27}. These datasets cover the head, chest, abdomen, and spine, as summarized in Fig.~\ref{fig:fewshot_overview}(a). DICOM and NIfTI data are processed separately. For DICOM data such as CQ500, raw intensities are converted into Hounsfield units (HU) using scanner-specific rescale slope and intercept parameters, whereas NIfTI images are loaded directly before the common preprocessing steps. All images are clipped to $[-1024,3072]$, resized to $512\times512$ using bilinear interpolation, normalized to $[0,1]$, and stored in H5 format. A total of 1,126,182 CT images are used for pre-training.

\subsubsection{Ultra-low-dose CT datasets}
Five datasets, including CQ500 (head) and four MSD tasks (Hepatic Vessel, Lung Tumours, Spleen, and Colon Cancer), are selected for ULDCT simulation, with normal-dose cases randomly sampled from each dataset. These five simulated datasets are used as the internal datasets. For external evaluation, KiTS19 and MSD Pancreas Tumours datasets are used to assess cross-organ generalization, while real experimentally acquired data are further included to evaluate robustness under practical acquisition conditions. To model organ-dependent acquisition variability, the incident photon flux follows the physics-based CT model in \cite{28}: \begin{equation} 
	Q_0(g) = K \cdot c \cdot mA(g) \cdot s 
\end{equation} 
where $Q_0(g)$ is the expected photon count, $g$ denotes the organ type, and $mA(g)$ is the organ-specific tube current obtained from imaging protocols or acquisition metadata. $K$, $c$, and $s$ denote the system-dependent scaling constant, collimation factor, and rotation time, respectively. For the 1/10 dose setting, the fixed photon counts are $1\times10^4$ for head, $1.2\times10^4$ for lung/colon, and $4\times10^3$ for spleen/hepatic vessel.

Normal-dose images are forward-projected to obtain line integrals $k$, followed by Poisson noise injection:
\begin{equation}
	I \sim \mathrm{Poisson}(N_0(g)\exp(-k))
\end{equation}
where $N_0(g)$ denotes the organ-specific photon count. The noisy measurements are converted back to the sinogram domain by
\begin{equation}
	k_{\mathrm{LD}} = -\log\left(\frac{I+\epsilon}{N_0(g)}\right)
\end{equation}
where $\epsilon$ avoids numerical instability. The noisy sinograms are reconstructed using FBP and normalized to $[0,1]$. Each organ contains 1000 training images, while the validation and test sets each contain 100 images per organ.

\subsection{Experimental setup}
\subsubsection{Pre-training details}
The DDN-based foundation model was pretrained on the curated multi-organ CT datasets. Full-image training was adopted, with Gaussian noise added online to each clean CT image. For blind-noise training, the noise level $\sigma$ was randomly sampled from $[0,75]$, and the noisy input was clipped to $[0,1]$. The model was optimized using Adam with $(\beta_1,\beta_2)=(0.9,0.99)$, an initial learning rate of $1\times10^{-4}$, and a total batch size of 8. The training objective combined a mean squared error denoising loss with an $\ell_1$ regularization term on the final sparse representation, with a weight of $1\times10^{-2}$. Automatic mixed precision was used. Pre-training was performed on $8\times$ NVIDIA GeForce RTX 4090 24\,GB GPUs.

\subsubsection{ULDCT denoising network training details}
For downstream tasks, a decoder module is attached to the pretrained DDN-based foundation model and jointly fine-tuned on multi-organ ULDCT datasets. The decoder maps the sparse representations produced by DDN back to the image space. The model is initialized with the pretrained DDN checkpoint, and all trainable parameters are jointly optimized from the beginning. Training is conducted for 100 epochs with a batch size of 1 on a single NVIDIA GeForce RTX 4090 24\,GB GPU. Adam is used with $(\beta_1,\beta_2)=(0.9,0.999)$ and an initial learning rate of $2\times10^{-4}$, which is reduced by a factor of 0.5 at the 40th and 80th epochs. Mean squared error is used as the training loss, together with automatic mixed precision. Peak signal-to-noise ratio (PSNR), structural similarity index measure (SSIM), and root mean square error (RMSE) are used for quantitative evaluation.

\subsection{Comparison with previous methods on internal datasets}
\label{formats}
To evaluate the denoising performance on the internal datasets, we compare DDN with FBP, LDCT denoising methods REDCNN \cite{3}, CNN10 \cite{4}, EDCNN \cite{5}, and FoundDiff \cite{21}, as well as all-in-one methods PromptIR \cite{17}, CAPTNet \cite{29}, MoCE-IR \cite{30}, and TAMP-S \cite{6}. DDN uses MoCE-Dec, i.e., the decoder module of MoCE-IR. All learning-based comparison methods are trained on the same five-organ mixed dataset using their officially released implementations for fair comparison.

\subsubsection{Quantitative results}
Table~\ref{tab:comparison} reports the quantitative results of different methods on the five organ test sets. DDN achieves the best average performance, with a PSNR of 42.99 dB, an SSIM of 0.9770, and an RMSE of 0.0072. Compared with MoCE-IR, the second-best method in terms of average PSNR and SSIM, DDN improves PSNR by 0.58 dB and SSIM by 0.0005, while reducing RMSE by 0.0005. DDN also achieves the best overall results on head, colon, lung, and spleen, and remains competitive on hepatic vessel. These results demonstrate that DDN can effectively accommodate anatomical differences and provide stable denoising performance across multiple organs.

\subsubsection{Qualitative results}
Fig.~\ref{fig:visual_indomain} presents visual comparisons on a representative lung case. FBP contains severe noise and loses local structural contrast. REDCNN, CNN10, and EDCNN suppress part of the noise, but residual artifacts and blurred tissue boundaries remain in the enlarged regions. FoundDiff and the all-in-one methods produce cleaner results, although some anatomical structures in the enlarged regions are over-smoothed or weakened. In comparison, DDN more effectively suppresses noise while preserving clear boundaries and subtle anatomical details.
\begin{figure*}[!t]
	\centering
	\includegraphics[
	width=\textwidth
	]{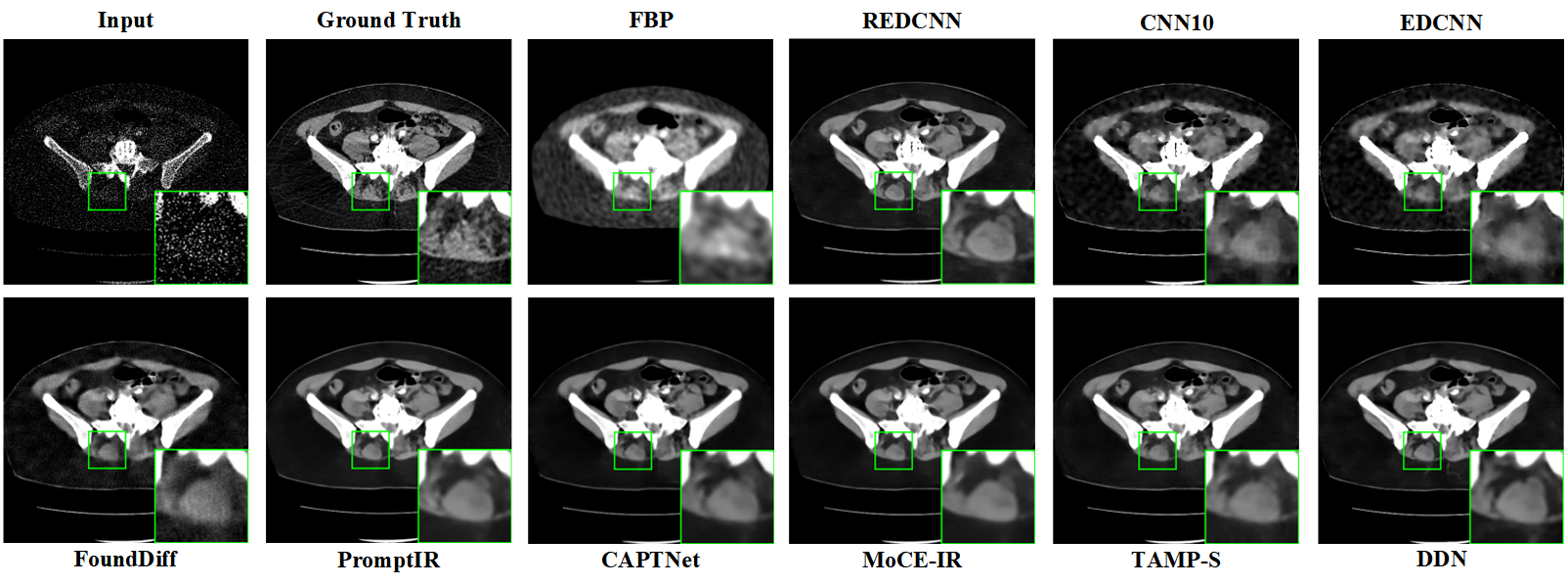}
	\caption{Visual comparison of different ULDCT denoising methods on a representative kidney case from the external datasets. Images are displayed using an HU window of [40, 350]. Green boxes indicate enlarged regions for detailed comparison of cross-organ denoising performance.}
	\label{fig:visual_ood}
\end{figure*}

\begin{figure*}[!t]
	\centering
	\includegraphics[width=\textwidth]{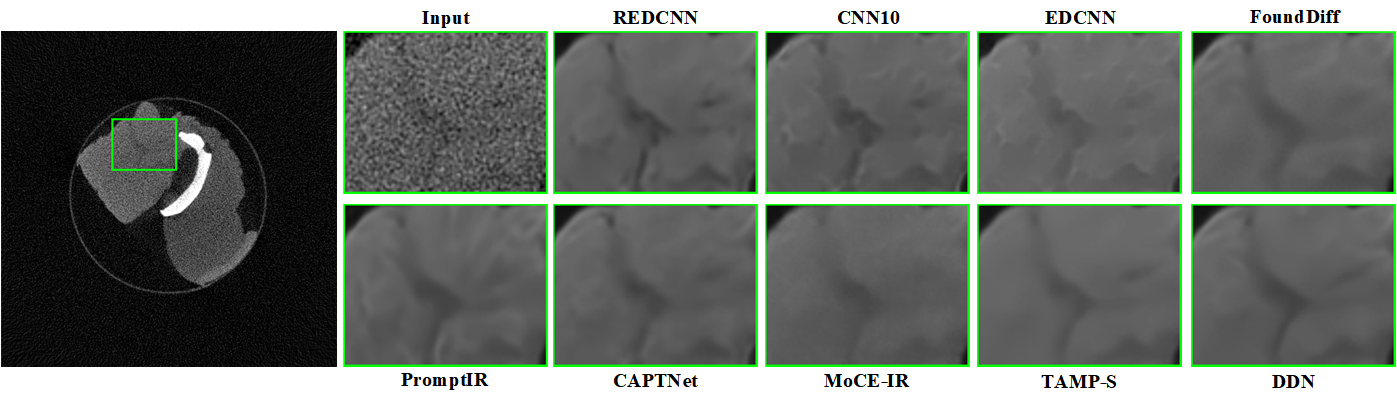}
	\caption{Visual comparison of different ULDCT denoising methods on real pork with bone data. Green boxes indicate enlarged regions for detailed comparison of noise suppression and structural preservation under practical acquisition conditions.}
	\label{fig:real_exp}
\end{figure*}

\subsection{Generalization on external datasets and real experimental data}
\label{sec}
\subsubsection{Generalization on external datasets}
To evaluate generalization on external datasets, all methods are directly tested on the KiTS19 (kidney) and Pancreas Tumours datasets. As shown in Table~\ref{tab:comparison_ood}, DDN achieves the best average performance, with a PSNR of 42.87 dB and an SSIM of 0.9795, outperforming PromptIR by 0.30 dB and 0.0004, respectively. The visual comparison in Fig.~\ref{fig:visual_ood} further shows that DDN more effectively suppresses residual noise while preserving clearer boundaries and internal anatomical details. These results demonstrate the strong cross-organ generalization of the proposed DDN.

\begin{table}[!t]
	\begingroup
	\centering
	\caption[Results on external datasets for kidney and pancreas.]%
	{\textnormal{Results on external datasets for kidney and pancreas. We report the average [PSNR (dB)$\uparrow$/SSIM$\uparrow$] values for each organ. The best and second-best results are highlighted in \capred{red} and \capblue{blue}, respectively.}}
	\label{tab:comparison_ood}
	
	\begin{tabular*}{\columnwidth}
		{@{\extracolsep{\fill}}lccc@{}}
		\toprule
		Method & Kidney & Pancreas & Average \\
		\midrule
		
		\multicolumn{4}{c}{\textnormal{LDCT Models}} \\
		\midrule
		FBP	& 30.47/0.8964& 31.11/0.8776& 30.79/0.8870 \\
		REDCNN& 41.14/0.9733& 42.48/0.9209& 41.81/0.9471 \\
		CNN10& 41.01/0.9673& 42.21/0.9671& 41.61/0.9672 \\
		EDCNN& 40.01/0.9576& 41.64/0.9613& 40.83/0.9595 \\
		FoundDiff& 40.93/0.9584& 41.93/0.8653& 41.43/0.9119 \\
		
		\midrule
		\multicolumn{4}{c}{\textnormal{All-in-one Models}} \\
		\midrule
		
		PromptIR
		& \textcolor{blue}{41.26}/0.9779
		& \textcolor{blue}{43.87}/\textcolor{red}{0.9802}
		& \textcolor{blue}{42.57}/\textcolor{blue}{0.9791} \\
		CAPTNet& 41.03/0.9780& 41.65/0.8633& 41.34/0.9207 \\
		MoCE-IR& 41.25/\textcolor{blue}{0.9788}	& 42.92/0.9257& 42.09/0.9523 \\
		TAMP-S& 40.91/0.9769& 43.20/0.9475& 42.06/0.9622 \\
		\textbf{DDN}
		& \textcolor{red}{41.31}/\textcolor{red}{0.9797}
		& \textcolor{red}{44.42}/\textcolor{blue}{0.9792}
		& \textcolor{red}{42.87}/\textcolor{red}{0.9795} \\
		
		\bottomrule
	\end{tabular*}
	
	\endgroup
	\normalcolor
\end{table}
\normalcolor

\subsubsection{Generalization on real experimental data}
\normalcolor
To evaluate the robustness of the proposed method under practical acquisition conditions, we conduct a comparison on real experimental data, including an equivalent water-bone phantom and pork with bone slices. Fig.~\ref{fig:real_exp} presents the visual result on the pork with bone image. From the enlarged regions, it can be observed that existing methods reduce noise to varying degrees, but residual noise, over-smoothing, and loss of subtle structures remain, particularly around low-contrast boundaries. In contrast, DDN more effectively suppresses noise while preserving structural integrity and fine local details, producing a visually cleaner result with better structural fidelity. These observations demonstrate the robustness and structural preservation capability of DDN on real experimental data.

\subsection{Effect of decoder design and foundation model}
To analyze the effects of the DDN-based foundation model and downstream decoder design, we compare three decoder settings: Conv41, Conv41 with two additional dictionary layers, and the decoder module of MoCE-IR, denoted as MoCE-Dec. Conv41 is a single convolutional decoder with a kernel size of $41\times41$, while MoCE-Dec adopts only the decoder module of MoCE-IR \cite{30}. Each decoder is evaluated with and without initialization from the pretrained DDN foundation model.

\begin{figure}[!t]
	\centering
	\includegraphics[
	width=\columnwidth,
	]{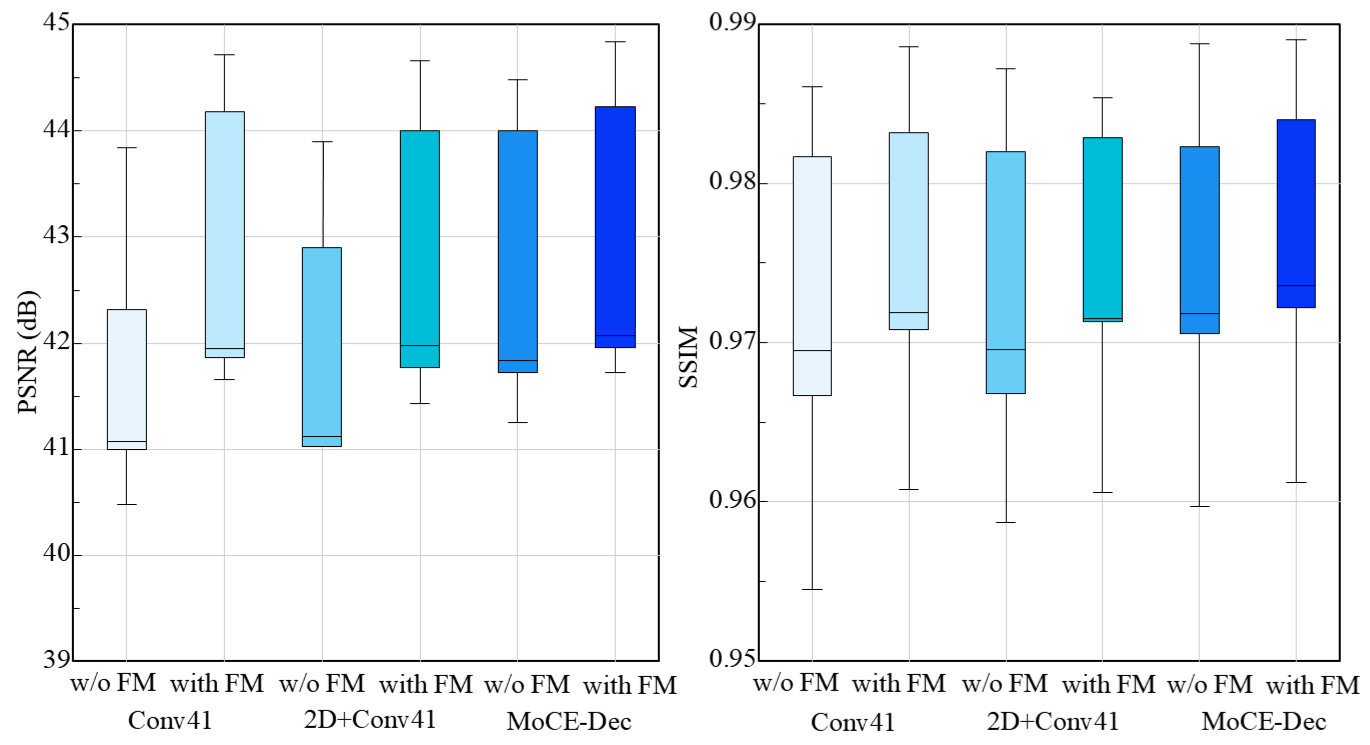}
	\caption{Effect of decoder design and foundation model initialization on ULDCT denoising. The boxplots show the PSNR and SSIM distributions over the five organ test sets. "2D+Conv41" denotes Conv41 with two additional dictionary layers, while "w/o FM" and "with FM" denote training without and with the pretrained DDN foundation model, respectively.}
	\label{fig:decoder_fm_effect}
\end{figure}

As shown in Fig.~\ref{fig:decoder_fm_effect}, the pretrained DDN foundation model consistently improves PSNR and SSIM across all decoder settings, indicating that its benefit is not limited to a specific decoder structure. Decoder design also affects the final performance, with MoCE-Dec producing the most favorable distributions. The combination of MoCE-Dec and the pretrained DDN achieves the best overall results, confirming the contributions of both foundation model pre-training and decoder design to downstream multi-organ ULDCT denoising.

\subsection{Few-shot learning}
To evaluate the effectiveness of the pretrained DDN under limited training data, we conduct few-shot experiments using only 20 training images for each organ. The validation and test sets remain unchanged, and all learning-based methods are retrained under the same setting.

\subsubsection{Comparison with previous methods} 
\begin{figure}[!t]
	\centering
	\includegraphics[
	width=\columnwidth,
	]{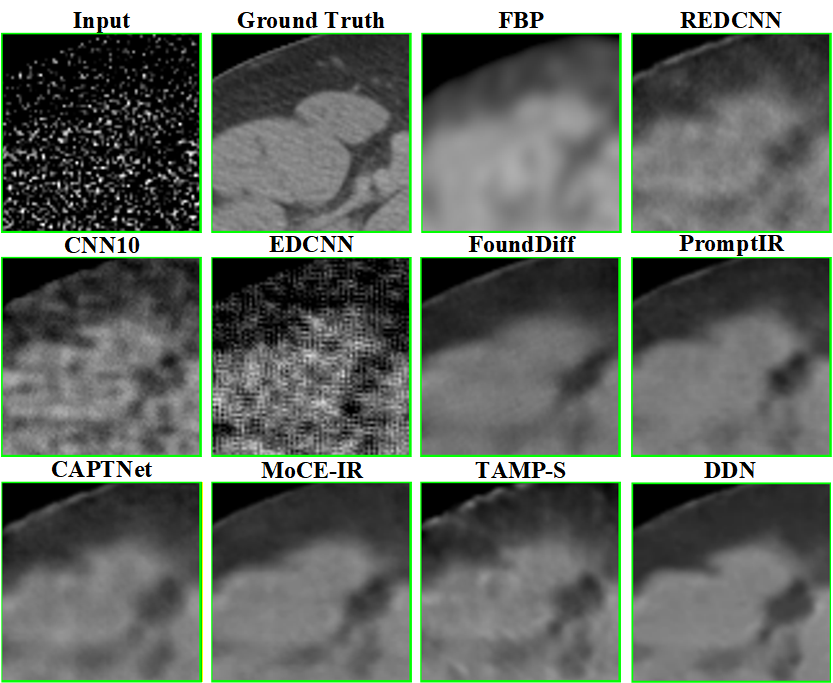}
	\caption{Visual comparison of different ULDCT denoising methods on a representative colon case under the few-shot setting. The enlarged regions are used for detailed comparison of local anatomical structures.}
	\label{fig:fig7}
\end{figure}

As shown in Fig.~\ref{fig:fewshot_overview}(b), DDN achieves the highest PSNR and SSIM among the compared methods, demonstrating superior few-shot denoising performance. Compared with MoCE-IR, DDN improves the average PSNR by 2.02 dB and SSIM by 0.0138, demonstrating that the pretrained sparse representations provide effective restoration priors when downstream data are limited. Figure~\ref{fig:fig7} presents the corresponding visual comparison. Existing methods reduce noise to varying degrees, but residual artifacts, over-smoothing, and weakened local boundaries remain around the highlighted region. In contrast, DDN produces a cleaner result while better preserving the subtle anatomical structure, further confirming the benefit of foundation model pre-training under the few-shot setting.

\subsubsection{Effect of decoder design and foundation model}

\begin{figure}[t]
	\centering
	\includegraphics[
	width=\columnwidth,
	]{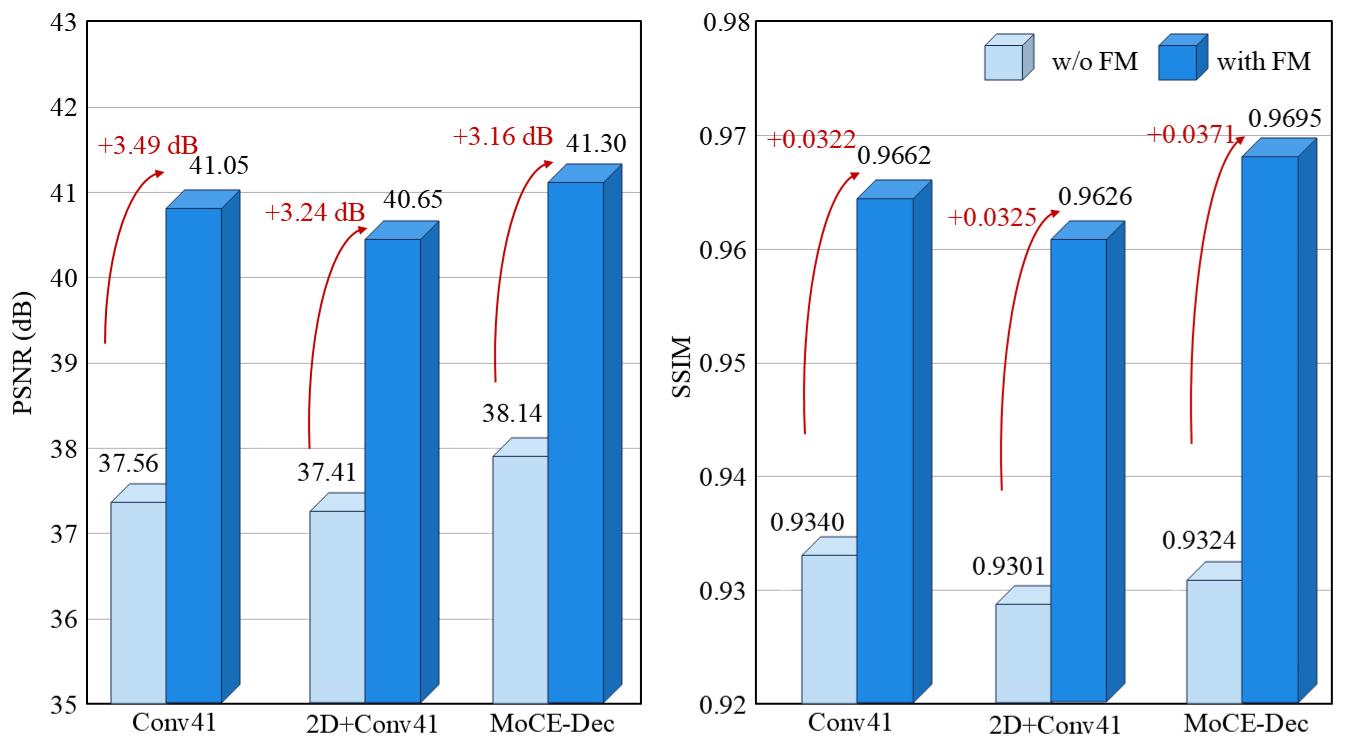}
	\caption{Effect of decoder design and foundation model initialization under the few-shot setting. PSNR and SSIM are reported for three decoder configurations with and without foundation model initialization, while the red annotations indicate the corresponding performance gains introduced by FM.}
	\label{fig:fig8}
\end{figure}
As shown in Fig.~\ref{fig:fig8}, initializing the model with the pretrained DDN produces clear improvements in both PSNR and SSIM across the three tested decoder configurations, with overall gains of approximately 3 dB in PSNR and 0.03 in SSIM. This consistent upward trend indicates that the pretrained DDN provides effective restoration priors under limited training data. Decoder design also affects the final performance. MoCE-Dec combined with foundation model initialization achieves the best average PSNR of 41.30 dB and SSIM of 0.9695, confirming the complementary benefits of pretrained restoration priors and an effective decoder for few-shot ULDCT denoising.

\subsection{Ablation studies}
To evaluate the contributions of the main components, we construct several variants by changing the dictionary depth and removing the DDM or individual sub-modules of the TGM. The results averaged over the five organ test sets are reported in Table~\ref{tab:ablation}.

As shown in Table~\ref{tab:ablation}, increasing the dictionary depth progressively improves denoising performance, with the complete three-layer DDN achieving the best results. This confirms that cascaded dictionary layers provide stronger capability for refining hierarchical sparse representations. Replacing the DDM with static convolutional kernels decreases the PSNR from 42.99 dB to 42.23 dB and degrades both SSIM and RMSE, demonstrating the importance of adapting the dictionary atoms to the input content. Removing any sub-module of TGM also leads to performance degradation, confirming that local, large, non-local, and global feature information contribute complementary cues to threshold generation. In particular, removing the non-local sub-module causes the largest decrease, reducing the PSNR by 1.72 dB, which indicates that long-range dependencies are important for preserving structural consistency. Overall, the full model achieves the best performance across all metrics, validating the combined effectiveness of the multilayer dictionary structure, DDM, and TGM.
\begin{table}[!t]
	\begingroup
	\centering
	\caption{\textnormal{Ablation studies on key components of the proposed method. The reported metrics are averaged over the test datasets. The best results are shown in bold.}}
	\label{tab:ablation}
	\begin{tabular*}{\columnwidth}
		{@{\extracolsep{\fill}}lccc@{}}
		\toprule
		\textbf{Method} & \textbf{PSNR (dB)} & \textbf{SSIM} & \textbf{RMSE} \\
		\midrule
		DDN w/ 1 dictionary layer
		& 42.49 & 0.9752 & 0.0076 \\
		DDN w/ 2 dictionary layers
		& 42.75 & 0.9762 & 0.0074 \\
		w/o local in TGM
		& 42.83 & 0.9768 & 0.0073 \\
		w/o large in TGM
		& 42.80 & 0.9762 & 0.0073 \\
		w/o non-local in TGM
		& 41.27 & 0.9694 & 0.0088 \\
		w/o global in TGM
		& 42.75 & 0.9762 & 0.0074 \\
		w/o DDM
		& 42.23 & 0.9740 & 0.0078 \\
		\textbf{DDN}
		& \textbf{42.99}
		& \textbf{0.9770}
		& \textbf{0.0072} \\
		
		\bottomrule
	\end{tabular*}	
	\endgroup
\end{table}

\section{Conclusion}
\label{sec:conclusion}

We proposed an architecture-interpretable foundation model for unified multi-organ ULDCT denoising. By combining multilayer convolutional sparse coding, DDM, TGM, and large-scale denoising pre-training with sparse regularization, DDN could learn transferable priors for multi-organ ULDCT denoising. Experiments on multiple organs, cross-organ generalization, real experimental data, and few-shot settings demonstrated its superior denoising performance and generalization capability. Future work will explore extending the DDN-based framework to other CT restoration tasks, such as artifact reduction and super-resolution.

\end{document}